%% file: S5PowerFlux1.tex
\documentclass[aps,prd,showpacs,amssymb,amsmath,amsfonts,superscriptaddress,
twocolumn,floatfix,preprintnumbers,altaffilletter]{revtex4}
\usepackage{graphicx}
\usepackage{hyperref}
\usepackage{amssymb}
\usepackage{amsmath}
\usepackage{longtable}
\usepackage{rotating}
\usepackage{color}
\usepackage{epsfig}
\usepackage{epsf}
\usepackage{fancyhdr}

\newcommand{\ee}[1]{\!\times\!10^{#1}}

\input defs.tex

\begin{document}
\pagestyle{fancy}
\rhead[]{}
\lhead[]{}
\title{ 
 All-sky LIGO Search for Periodic Gravitational Waves in the Early S5 Data }
\input{authorlist}

\date{\today}

\begin{abstract}
  We report on an all-sky search with the LIGO detectors for 
  periodic gravitational waves in the frequency   range $50\,$--$\,1100$~Hz and   
  with the frequency's time derivative in the range $-\sci{5}{-9}$--$\,0$~Hz s$^{-1}$.
  Data from the first eight months of the fifth LIGO science run (S5) have been 
  used in this search, which is based on a semi-coherent method (PowerFlux) of 
  summing strain power. Observing no evidence
  of periodic gravitational radiation, we report 95\%\ confidence-level 
  upper limits on radiation
  emitted by any unknown isolated rotating neutron stars
  within the search range. Strain limits below $10^{-24}$ are obtained over
  a 200-Hz band, and the sensitivity improvement over previous searches increases the spatial volume 
  sampled by an average factor of about 100 over the entire search band. 
  For a neutron star with nominal equatorial ellipticity of
  $10^{-6}$, the search is sensitive to distances as great as 500 pc---a
  range that could encompass many undiscovered neutron stars, albeit
  only a tiny fraction of which would likely be rotating fast enough to be accessible to LIGO.
  This ellipticity is at the upper range thought to be sustainable by conventional
  neutron stars and well below the maximum sustainable by a strange quark star.

\end{abstract}
\pacs{04.80.Nn, 95.55.Ym, 97.60.Gb, 07.05.Kf}
\preprint{LIGO-P080024-03-Z}
\maketitle
%
\section{Introduction}
\label{sec:introduction}

We have carried out an all-sky search with the LIGO (Laser Interferometer
Gravitational-wave Observatory) detectors~\cite{ligo1,ligo2} for
periodic gravitational waves, using data from the first eight months 
of LIGO's fifth science run (S5). We have searched over the frequency range $50\,$--$\,1100$~Hz,
allowing for a frequency time derivative in the range $-\sci{5}{-9}$--$\,0$~Hz s$^{-1}$.
Rotating neutron stars in our galaxy are the prime target. At signal frequencies near 100 Hz
we obtain strain sensitivities below $10^{-24}$, a strain at
which one might optimistically expect to see the strongest
signal from a previously unknown neutron star according to a generic
argument originally made by Blandford (unpublished), and extended in our
previous search for such objects in S2 data~\cite{S2FstatPaper}. 
A recent refinement of the argument~\cite{knispelallen} gives 
less optimistic estimates, but these too are surpassed by the experimental
results presented here. 

Using data from earlier science runs, the LIGO Scientific Collaboration (LSC) has previously
reported on 
all-sky searches for unknown
rotating neutron stars (henceforth designated as ``pulsars'' here).
These searches have been performed using a short-period coherent search in the $160.0\,$--$\,728.8$~Hz
frequency range~\cite{S2FstatPaper}, and using a long-period semi-coherent search in
the $200\,$--$\,400$~Hz frequency range in the S2 data~\cite{S2HoughPaper} and the
$50\,$--$\,1000$~Hz range in the S4 data~\cite{S4IncoherentPaper}.
Einstein@Home, a distributed home computing effort~\cite{BOINC},
has also been running searches using a coherent first stage, followed by a simple coincidence
stage, for which S3 and S4 results have been released~\cite{S3EatH,S4EatH}.

The data collected in the S5 data run were more sensitive than in previous data runs,
and the amount of data used here is an increase by a factor of eight over that reported
from the S4 data run~\cite{S4IncoherentPaper}, resulting in upper limits on periodic
gravitational waves about a factor of $3\,$--$\,6$ lower than those from the S4 data, depending
on source frequency. This improvement gives an increase in sampled galactic volume by about
a factor of 100, depending on the assumed source frequency
and spin-down. At a signal frequency of 1100 Hz 
we achieve sensitivity to neutron stars of equatorial ellipticity $\epsilon\sim10^{-6}$
at distances up to 500 pc (see \cite{S4IncoherentPaper} for relations). 
This ellipticity is at the upper range thought to be sustainable by conventional 
neutron stars~\cite{ushomirskyetal}  and well below the maximum 
sustainable (10$^{-4}$) by a strange quark star~\cite{owenstrangestar}.
The number of undiscovered, electromagnetically quiet neutron stars 
within 500 pc can be estimated to be $O(10^4-10^5)$ from the 
neutron star birth rate~\cite{narayan}, although it is likely that only a tiny fraction would both
be rotating fast enough to be accessible to LIGO~\cite{lorimer} and remain in
the local volume over the age of the galaxy~\cite{cordeschernoff}. Only 
$\sim25$ radio or x-ray pulsars have been discovered so far within that volume~\cite{ATNF}.

\section{The LIGO Detectors and the S5 Science Run}
\label{sec:detectordata}

The LIGO detector network consists of a 4-km interferometer in
Livingston Louisiana, (L1), and two interferometers in Hanford
Washington, one 4-km and the other 2-km (H1 and H2). 

The data analyzed in this paper were produced in the first eight months of
LIGO's fifth science run (S5). This run started at 16:00 UTC on November 4, 2005
at the LIGO Hanford Observatory and at 16:00 UTC on November 14, 2005 at the
LIGO Livingston Observatory; the run ended at 00:00 UTC on October 1, 2007.
During this run, all three LIGO
detectors had displacement spectral amplitudes very near their design
goals of $\sci{1.1}{-19} {\rm m} \, {\rm Hz}^{-1/2}$~\cite{S5DetectorPaper}
in their most sensitive frequency band near 150 Hz.
(In terms of gravitational-wave strain, the H2 interferometer was
roughly a factor of two less sensitive than
the other two; its data were not used in this search.)

The data were acquired and digitized at a rate of 16384~Hz.
Data acquisition was periodically interrupted by disturbances
such as seismic transients (natural or anthropogenic),
reducing the net running time of the interferometers.
In addition, there were 1--2 week commissioning breaks to
repair equipment and address newly identified noise sources.
The resulting duty factors for the interferometers 
were approximately 69\% for H1, 77\% for H2, and 
57\% for L1 during the first eight months. A nearby construction
project degraded the L1 duty factor significantly during this
early period of the S5 run. 
By the end of the S5 run, the cumulative duty factors had improved
to 78\% for H1, 79\% for H2, and 66\% for L1. For this search,
approximately 4077 hours of H1 data and 3070 hours of L1 data were used, where
each data segment used was required to contain at least 30 minutes
of continuous interferometer operation.

\section{Signal Waveforms}
\label{sec:waveforms}

The general form of a gravitational-wave signal is described in terms of
two orthogonal transverse polarizations defined to be ``$+$'' with waveform $h_+(t)$ 
and ``$\times$'' with waveform $h_\times(t)$, for which
separate and time-dependent antenna pattern factors $F_+$ and $F_\times$ apply,
which depend on a polarization angle $\psi$~\cite{jks}.
For periodic gravitational waves, which in general are elliptically polarized,
the individual components $h_{+,\times}$ have the form
$h_+(t) = \hpluszero \cos\Phi(t)$ and $h_\times(t) = \hcrosszero \sin\Phi(t)$,
where $\hpluszero$ and $\hcrosszero$ are the amplitudes
of the two polarizations, and $\Phi(t)$ is the phase of the signal at the detector.
For the semi-coherent method used in this search, only the
instantaneous signal frequency in the detector
reference frame, $2\pi f(t) = d\Phi(t)/dt$, needs to be calculated. 
For an isolated, precession-free, rigidly rotating neutron star
the quadrupolar amplitudes $\hpluszero$ and $\hcrosszero$ are related 
to wave amplitude, $h_0$, by $\hpluszero = h_0 \frac{1+ \cos^2\iota}{2}$
and $\hcrosszero = h_0 \cos\iota$,
where $\iota$ describes the inclination angle of the star's 
spin axis with respect to the line of sight. For such a star,
the signal wave frequency, $f$, is twice the rotation frequency, $f_r$.

The detector reference frame frequency $f(t)$
can, to a very good approximation, be related
to the frequency  $\fhat(t)$ in the Solar System Barycenter (SSB) frame by~\cite{S2HoughPaper}
$f(t) - \hat{f}(t) \simeq \hat{f}(t)\frac{ {\bf v} (t)\cdot\bf{n}}{c}$,
where ${\bf v}(t)$ is the detector's velocity with respect to the SSB frame,
and $\bf{n}$ is the unit-vector pointing from the detector toward the sky location of
the source~\cite{S2HoughPaper}. 

\section{Analysis Method}
\label{sec:analysismethodoverview}

The PowerFlux method used in this analysis is described in detail
elsewhere~\cite{S4IncoherentPaper} and is a variation upon the
StackSlide method~\cite{BC00}. Here we summarize briefly its
main features. 

A strain power estimator is derived from summing measures of strain power from
many short, 50\%-overlap, Hann-windowed Fourier transforms (SFTs) that have been created
from 30-minute intervals of calibrated strain data.  In searching a narrow
frequency range (0.5~mHz spacing) for an assumed source sky location,
explicit corrections are made for Doppler modulations of the apparent source
frequency.  These modulations are due to the Earth's rotation and its orbital motion around
the SSB, and the frequency's time derivative, $\dot{f}$, intrinsic to
the source. Corrections are also applied for antenna pattern modulation,
assuming five different polarizations: four
linear polarizations separated by $\pi/8$ in polarization angle, and
circular polarization. When summing, the variability of the noise
is taken into account with an SFT-dependent weight proportional to
the expected inverse variance of the background noise power (see
\cite{S4IncoherentPaper,PowerFluxTechNote} for detailed formulae).

The search range for initial frequency $\fhatzero$ values is 
$50\,$--$\,1100$~Hz with a uniform grid spacing equal to the size of an SFT frequency bin [1/(30 min)].
The range of $\dot{f}$ values searched is $-\sci{5}{-9}$--$\,0$~Hz s$^{-1}$ with a spacing
of $\sci{5}{-10}$~Hz s$^{-1}$,
since isolated rotating neutron stars are generally expected to spin down with time.
As discussed in our previous reports~\cite{S2HoughPaper, S2FstatPaper, S4IncoherentPaper}, 
the number of sky points
that must be searched grows quadratically with
the frequency $\fhatzero$, ranging here from about five thousand at
50~Hz to about 2.4 million at 1100~Hz.  The sky grid used here is
isotropic and covers the entire sky. 

Upper limits calculated in this method are strict frequentist limits on linear
and circular polarization in small patches on the sky, with the limits
quoted here being the highest limits in each 0.25-Hz band over broad regions of the  sky.
These are interpreted as limits on worst-case (linear polarization) and 
best-case (circular polarization) orientations of rotating neutron stars.
Since the eight
months of data analyzed here cover a large span of the Earth's orbit, providing
substantial Doppler modulation of source frequency,
contamination from stationary instrumental lines is much reduced from 
earlier and shorter data runs. A total of only
0.6\% of the search volume in sky location and spindown had to be excluded
from the upper-limit analysis because of Doppler stationarity.

The primary changes in the PowerFlux algorithm used in this search concern
followup of outlier candidates. (The general method for setting upper limits is identical
to that used in the S4 search~\cite{S4IncoherentPaper}.) 
Here we summarize the followup method used. Single-interferometer searches are carried out
separately for the H1 and L1 interferometers, leading to the upper limits on strain
shown in Fig.~\ref{fig:upperlimits} and discussed below.  
During determination of the maximum upper limit per sky region, per frequency band 
and per spin-down step, a ``domain map'' is constructed of local signal-to-noise ratio (SNR) maxima,
with the domains ordered by maximum gridpoint SNR and clustered if close in direction and frequency.
The 1000 domains with the highest maximum SNR are then re-analyzed to
obtain improved estimates of the associated candidate parameters,
using a modified
gradient search with a matched filter to maximize SNR with respect to source frequency, spin-down,
sky location, polarization angle $\psi$, and inclination angle $\iota$~\cite{PowerFluxTechNote}.
This maximization step samples frequency and spin-down much more
finely than in the initial search.

When all sky regions and all spin-downs have been searched for a given
0.25~Hz band for both H1 and L1, the search pipeline outputs are compared,
and the following criteria are used to define candidates
for followup analysis. The H1 and L1 candidates must
each have an SNR value greater than 6.25, and they must
agree in frequency to within 1/180~Hz = 5.56~mHz, in spin-down 
to within 4$\times$10$^{-10}\textrm{~Hz}\textrm{~s}^{-1}$, and in sky location to within 0.14 radians.
These conservative choices have been guided by simulated single-interferometer pulsar injections.
Coincidence candidates within 0.1 Hz of one another are grouped together,
since most candidates arise from detector spectral artifacts that become apparent upon
manual investigation.

Candidates passing these criteria are subjected to a computationally intensive 
followup analysis that reproduces the all-sky PowerFlux search in a
0.25 Hz band around each candidate, this time using the (incoherently)
combined strain powers from both interferometers.
Sky maps of strain and SNR are created and examined manually
for each individual interferometer and for the combined interferometers. 
Spectral estimates from noise decomposition are also examined to identify possible
artifacts leading to the coincident outliers. 

\section{Results}
\label{sec:results}

\begin{figure}
  \begin{center}
  \includegraphics[height=8.0cm]{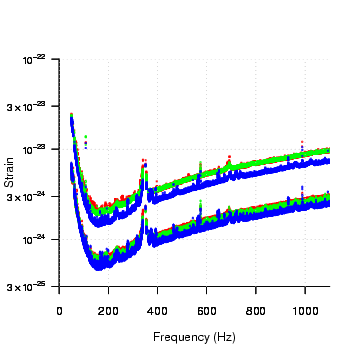}
  \caption{Minimum (H1 or L1) upper limits (95\%\ CL) on pulsar gravitational wave
  amplitude $h_0$ for the equatorial (red), intermediate (green), and polar (blue) 
  declination bands for best-case (lower curves) and worst-case (upper curves)
  pulsar orientations. Shown are all the minimum limits for each of the 11
  spin-down values from $-\sci{5}{-9}\,\textrm{Hz}\,\textrm{s}^{-1}$ to zero in steps 
  of $\sci{5}{-10}\,\textrm{Hz}\,\textrm{s}^{-1}$.}
  \label{fig:upperlimits} 
  \end{center}
\end{figure}

Figure~\ref{fig:upperlimits}
shows the lower of the H1 and L1 95\%\ confidence-level upper limits
on pulsar gravitational wave amplitude $h_0$ 
for worst-case and best-case pulsar orientations for different declination
bands (each with different run-averaged antenna pattern sensitivity).
As in the S4 analysis, narrow bands around 60-Hz power mains harmonics,
along with bands characterized by non-Gaussian noise, have been excluded from the
displayed limits. Numerical values for frequencies and limits displayed 
in these figures can be obtained
separately~\cite{epaps}. Systematic uncertainties on these values are dominated by calibration
uncertainty at the $\sim$10\% level.

All outliers were checked for coincidence between H1 and L1, as described above.
In most cases single-interferometer spectral artifacts were readily found
upon initial inspection, most of which had known instrumental or environmental
sources, such as mechanical resonances (``violin modes'') of the wires supporting
interferometer mirrors, and power mains harmonics of 60 Hz. Other outliers
were tracked down to previously unknown electromagnetic disturbances.
For six coincidence candidates, no instrumental spectral artifacts were 
apparent. Their 0.1-Hz bands and favored spin-down values are listed
in Table~\ref{tab:candidateshard}, along with the maximum SNR's observed in H1 and L1
data. 

None of these six candidates was confirmed, however, 
as a detection of a constant-amplitude, constant-spin-down
periodic source of gravitational radiation. 
In each case, we found that the combined H1-L1 SNR did not increase by
more than 0.6 (0.4) units over the minimum (maximum) of the
single-interferometer SNR's, with four candidates showing a
{\it decrease} for combined SNR.
To understand the expectation for a true signal, we carried
out {\it a posteriori} software signal injections, which indicated that combined SNR should
typically show an increase over minimum SNR by more than
2.0 units for a single-interferometer SNR threshold of 6.25.
Hence we conservatively veto all candidates with
an SNR increase less than 1 unit. In addition, manual exploration of these
candidates was carried out, using larger portions of the S5 run's
data, to determine whether SNR increased with additional data,
and with subsets of the the original 8-month data, to determine
whether a transient astrophysical source could explain the candidate.
None of these explorations proved fruitful.

We also note that multi-interferometer injections indicate that for signal
frequencies above 850 Hz, the coincidence 
requirements in frequency and sky location could be tightened by a factor of 
five to 1 mHz and by a factor of seven to 0.02 radians, respectively, 
with only a slight reduction in efficiency for true signals. None of the candidates in
Table~\ref{tab:candidateshard} satisfies these tighter criteria.

\begin{table}
\begin{tabular}{ccccc}\hline
Frequency band  &   Spin-down   &        &       \\
     (Hz)       & (Hz s$^{-1}$) & H1 SNR & L1 SNR\\
\hline \hline
867.2 & $-4.3\times10^{-9}$ & 6.27 & 6.30  \\
941.0 & $-2.0\times10^{-9}$ & 6.50 & 6.67 \\
967.8 & $-1.5\times10^{-9}$ & 6.26 & 6.33  \\
979.5 & $-5.0\times10^{-9}$ & 6.40 & 6.29  \\
1058.6 & $-5.0\times10^{-10}$ & 6.83 & 6.38 \\
1070.2 & $-3.0\times10^{-10}$ & 6.72 & 6.99 \\
\hline
\end{tabular}
\caption{List of coincidence candidates for which no instrumental
         spectral artifacts were observed.}
\label{tab:candidateshard}
\end{table}

In summary, we have set strict, all-sky frequentist upper limits on the strength of continuous-wave
gravitational radiation of linear and circular polarization,
corresponding to least favorable
and most favorable pulsar orientations, respectively.
Followup analysis of coincidence candidates with SNR $>$ 6.25
did not yield a detection. The limits on detected strain can be
translated into limits on equatorial ellipticity as small as 
$10^{-6}$ for unknown neutron stars as far away as 500 pc.
This ellipticity is at the upper range thought to be sustainable by conventional 
neutron stars  and well below the maximum 
sustainable (10$^{-4}$) by a strange quark star.

\section{Acknowledgments}
\input acknowledgements.tex
This document  has been assigned LIGO Laboratory document number
LIGO-P080024-03-Z.


%

\end{document}

%% file: defs.tex
\def\bea{\begin{eqnarray}}
\def\eea{\end{eqnarray}}
\def\beq{\begin{equation}}
\def\eeq{\end{equation}}

\def\be{\begin{equation}}
\def\ee{\end{equation}}

\newcommand{\ba}{\begin{eqnarray}}
\newcommand{\ea}{\end{eqnarray}}

\newcommand{\bml}{\begin{mathletters}}
\newcommand{\eml}{\end{mathletters}}

\def\sci#1#2{#1\times10^{#2}}

\def\fhat{\hat f}
\def\hpluszero{A_{+}}
\def\hcrosszero{A_{\times}}
\def\fhatzero{\hat{f}_0}



%% file: authorlist.tex
\newcommand*{\AG}{Albert-Einstein-Institut, Max-Planck-Institut f\"ur Gravitationsphysik, D-14476 Golm, Germany}
\affiliation{\AG}
\newcommand*{\AH}{Albert-Einstein-Institut, Max-Planck-Institut f\"ur Gravitationsphysik, D-30167 Hannover, Germany}
\affiliation{\AH}
\newcommand*{\AU}{Andrews University, Berrien Springs, MI 49104 USA}
\affiliation{\AU}
\newcommand*{\AN}{Australian National University, Canberra, 0200, Australia}
\affiliation{\AN}
\newcommand*{\CH}{California Institute of Technology, Pasadena, CA  91125, USA}
\affiliation{\CH}
\newcommand*{\CA}{Caltech-CaRT, Pasadena, CA  91125, USA}
\affiliation{\CA}
\newcommand*{\CU}{Cardiff University, Cardiff, CF24 3AA, United Kingdom}
\affiliation{\CU}
\newcommand*{\CL}{Carleton College, Northfield, MN  55057, USA}
\affiliation{\CL}
\newcommand*{\CS}{Charles Sturt University, Wagga Wagga, NSW 2678, Australia}
\affiliation{\CS}
\newcommand*{\CO}{Columbia University, New York, NY  10027, USA}
\affiliation{\CO}
\newcommand*{\ER}{Embry-Riddle Aeronautical University, Prescott, AZ   86301 USA}
\affiliation{\ER}
\newcommand*{\HC}{Hobart and William Smith Colleges, Geneva, NY  14456, USA}
\affiliation{\HC}
\newcommand*{\IA}{Institute of Applied Physics, Nizhny Novgorod, 603950, Russia}
\affiliation{\IA}
\newcommand*{\IU}{Inter-University Centre for Astronomy  and Astrophysics, Pune - 411007, India}
\affiliation{\IU}
\newcommand*{\HU}{Leibniz Universit{\"a}t Hannover, D-30167 Hannover, Germany}
\affiliation{\HU}
\newcommand*{\CT}{LIGO - California Institute of Technology, Pasadena, CA  91125, USA}
\affiliation{\CT}
\newcommand*{\LM}{LIGO - Massachusetts Institute of Technology, Cambridge, MA 02139, USA}
\affiliation{\LM}
\newcommand*{\LO}{LIGO Hanford Observatory, Richland, WA  99352, USA}
\affiliation{\LO}
\newcommand*{\LV}{LIGO Livingston Observatory, Livingston, LA  70754, USA}
\affiliation{\LV}
\newcommand*{\LU}{Louisiana State University, Baton Rouge, LA  70803, USA}
\affiliation{\LU}
\newcommand*{\LE}{Louisiana Tech University, Ruston, LA  71272, USA}
\affiliation{\LE}
\newcommand*{\LL}{Loyola University, New Orleans, LA 70118, USA}
\affiliation{\LL}
\newcommand*{\MS}{Moscow State University, Moscow, 119992, Russia}
\affiliation{\MS}
\newcommand*{\ND}{NASA/Goddard Space Flight Center, Greenbelt, MD  20771, USA}
\affiliation{\ND}
\newcommand*{\NA}{National Astronomical Observatory of Japan, Tokyo  181-8588, Japan}
\affiliation{\NA}
\newcommand*{\NO}{Northwestern University, Evanston, IL  60208, USA}
\affiliation{\NO}
\newcommand*{\RA}{Rutherford Appleton Laboratory, Chilton, Didcot, Oxon OX11 0QX United Kingdom}
\affiliation{\RA}
\newcommand*{\SJ}{San Jose State University, San Jose, CA 95192, USA}
\affiliation{\SJ}
\newcommand*{\SM}{Sonoma State University, Rohnert Park, CA 94928, USA}
\affiliation{\SM}
\newcommand*{\SE}{Southeastern Louisiana University, Hammond, LA  70402, USA}
\affiliation{\SE}
\newcommand*{\SO}{Southern University and A\&M College, Baton Rouge, LA  70813, USA}
\affiliation{\SO}
\newcommand*{\SA}{Stanford University, Stanford, CA  94305, USA}
\affiliation{\SA}
\newcommand*{\SR}{Syracuse University, Syracuse, NY  13244, USA}
\affiliation{\SR}
\newcommand*{\PU}{The Pennsylvania State University, University Park, PA  16802, USA}
\affiliation{\PU}
\newcommand*{\TA}{The University of Texas at Austin, Austin, TX 78712, USA}
\affiliation{\TA}
\newcommand*{\TC}{The University of Texas at Brownsville and Texas Southmost College, Brownsville, TX  78520, USA}
\affiliation{\TC}
\newcommand*{\TR}{Trinity University, San Antonio, TX  78212, USA}
\affiliation{\TR}
\newcommand*{\BB}{Universitat de les Illes Balears, E-07122 Palma de Mallorca, Spain}
\affiliation{\BB}
\newcommand*{\UA}{University of Adelaide, Adelaide, SA 5005, Australia}
\affiliation{\UA}
\newcommand*{\BR}{University of Birmingham, Birmingham, B15 2TT, United Kingdom}
\affiliation{\BR}
\newcommand*{\FA}{University of Florida, Gainesville, FL  32611, USA}
\affiliation{\FA}
\newcommand*{\GU}{University of Glasgow, Glasgow, G12 8QQ, United Kingdom}
\affiliation{\GU}
\newcommand*{\MD}{University of Maryland, College Park, MD 20742 USA}
\affiliation{\MD}
\newcommand*{\MA}{University of Massachusetts, Amherst, MA 01003 USA}
\affiliation{\MA}
\newcommand*{\MU}{University of Michigan, Ann Arbor, MI  48109, USA}
\affiliation{\MU}
\newcommand*{\MN}{University of Minnesota, Minneapolis, MN 55455, USA}
\affiliation{\MN}
\newcommand*{\OU}{University of Oregon, Eugene, OR  97403, USA}
\affiliation{\OU}
\newcommand*{\RO}{University of Rochester, Rochester, NY  14627, USA}
\affiliation{\RO}
\newcommand*{\SL}{University of Salerno, 84084 Fisciano (Salerno), Italy}
\affiliation{\SL}
\newcommand*{\SN}{University of Sannio at Benevento, I-82100 Benevento, Italy}
\affiliation{\SN}
\newcommand*{\SH}{University of Southampton, Southampton, SO17 1BJ, United Kingdom}
\affiliation{\SH}
\newcommand*{\SC}{University of Strathclyde, Glasgow, G1 1XQ, United Kingdom}
\affiliation{\SC}
\newcommand*{\WA}{University of Western Australia, Crawley, WA 6009, Australia}
\affiliation{\WA}
\newcommand*{\UW}{University of Wisconsin-Milwaukee, Milwaukee, WI  53201, USA}
\affiliation{\UW}
\newcommand*{\WU}{Washington State University, Pullman, WA 99164, USA}
\affiliation{\WU}

\author{B.~Abbott}    \affiliation{\CT}
\author{R.~Abbott}    \affiliation{\CT}
\author{R.~Adhikari}    \affiliation{\CT}
\author{P.~Ajith}    \affiliation{\AH}
\author{B.~Allen}    \affiliation{\AH}  \affiliation{\UW}
\author{G.~Allen}    \affiliation{\SA}
\author{R.~Amin}    \affiliation{\LU}
\author{S.~B.~Anderson}    \affiliation{\CT}
\author{W.~G.~Anderson}    \affiliation{\UW}
\author{M.~A.~Arain}    \affiliation{\FA}
% ...

 \collaboration{The LIGO Scientific Collaboration, http://www.ligo.org}
 \noaffiliation

%% file: acknowledgements.tex
We thank Deepto Chakrabarty and David Kaplan for
useful discussions.
The authors gratefully acknowledge the support of the United States
National Science Foundation for the construction and operation of the
LIGO Laboratory and the Science and Technology Facilities Council of the
United Kingdom, the Max-Planck-Society, and the State of
Niedersachsen/Germany for support of the construction and operation of
the GEO600 detector. The authors also gratefully acknowledge the support
of the research by these agencies and by the Australian Research Council,
the Council of Scientific and Industrial Research of India, the Istituto
Nazionale di Fisica Nucleare of Italy, the Spanish Ministerio de
Educaci\'on y Ciencia, the Conselleria d'Economia Hisenda i Innovaci\'o of
the Govern de les Illes Balears, the Royal Society, the Scottish Funding 
Council, the Scottish Universities Physics Alliance, The National Aeronautics
and Space Administration, the Carnegie Trust, the Leverhulme Trust, the David
and Lucile Packard Foundation, the Research Corporation, and the Alfred
P. Sloan Foundation. 